\documentclass[reprint,prb,showpacs,preprintnumbers,amsmath,amssymb,twocolumns,footinbib,citeautoscript]{revtex4-1}

\usepackage{graphicx}
\usepackage{dcolumn}
\usepackage{bm}

\begin{document}

\title{Theoretical study of the charge transport through C$_{60}$-based 
single-molecule junctions}

\author{S. Bilan,$^1$ L. A. Zotti,$^1$  F. Pauly,$^{2,3}$ and J. C. Cuevas$^{1}$}

\affiliation{$^1$Departamento de F\'{\i}sica Te\'orica de la Materia Condensada,
Universidad Aut\'onoma de Madrid, E-28049 Madrid, Spain \\
$^2$Institut f\"ur Theoretische Festk\"orperphysik and DFG Center for Functional 
Nanostructures, Karlsruhe Institute of Technology, D-76131 Karlsruhe, Germany \\ 
$^3$Molecular Foundry, Lawrence Berkeley National Laboratory, Berkeley, California 
94720, USA}

\date{\today}

\begin{abstract}
We present a theoretical study of the conductance and thermopower of single-molecule
junctions based on C$_{60}$ and C$_{60}$-terminated molecules. We first analyze
the transport properties of gold-C$_{60}$-gold junctions and show that these junctions 
can be highly conductive (with conductances above $0.1G_0$, where $G_0=2e^2/h$ is the quantum 
of conductance). Moreover, we find that the thermopower in these junctions is negative 
due to the fact that the lowest unoccupied molecular orbital (LUMO) dominates the charge 
transport, and its magnitude can reach several tens of $\mu$V/K, depending on the contact 
geometry. On the other hand, we study the suitability of C$_{60}$ as an anchoring group 
in single-molecule junctions. For this purpose, we analyze the transport through several 
dumbbell derivatives using C$_{60}$ as anchors, and we compare the results with those 
obtained with thiol and amine groups. Our results show that the conductance of 
C$_{60}$-terminated molecules is rather sensitive to the binding geometry. Moreover, the 
conductance of the molecules is typically reduced by the presence of the C$_{60}$ anchors,
which in turn makes the junctions more sensitive to the functionalization of the 
molecular core with appropriate side groups. 
\end{abstract}


\maketitle

\section{Introduction}
The fullerene C$_{60}$ is attracting a lot of attention in the field of molecular 
electronics.\cite{Cuevas2010} One reason is that the delocalization of
the frontier orbitals of C$_{60}$ suggests that this molecule can be a good 
candidate to build highly conductive single-molecule junctions, a goal that 
remains elusive and that it has only being achieved with short molecules.\cite{Smit2002,
Kiguchi2008,Cheng2011,Chen2011,Kim2011} On the other hand, it has been 
recently suggested that C$_{60}$ used as an anchoring group to bind molecules 
to the electrodes can improve the reproducibility of the conductance
measurements in single-molecule junctions,\cite{Martin2008} which is a crucial 
issue in molecular electronics. The goal of this work is to further analyze these 
two different questions from a theoretical point of view.

The first experiment on individual C$_{60}$ molecules was reported by Joachim
\emph{et al.}\cite{Joachim1995} Here, a scanning tunneling microscope (STM), with
a tip made of tungsten, was used at room temperature to study the conductance of a 
C$_{60}$ molecule on an Au(110) surface. It was shown that in the contact regime
this heterojunction has a conductance of $2.35 \times 10^{-4}G_0$, which is
clearly lower than in the case of metallic atomic-size contacts.
Since then, different groups have investigated experimentally the transport 
properties of C$_{60}$ molecular junctions, mainly with gold electrodes, and
they have reported very different results. Thus for instance, Park 
\emph{et al.}\cite{Park2000} performed measurements in Au-C$_{60}$-Au junctions
using the electromigration technique and depositing the C$_{60}$ molecules from a 
liquid solution. In this case, the conductance at low bias was found to be largely 
suppressed and the current-voltage characteristics were dominated by the Coulomb 
blockade phenomenon. A related experiment, but this time with a microfabricated 
break junction, showed a much higher low-bias conductance (of the order of 
$0.1G_0$), which was attributed to the appearance of Kondo physics.\cite{Parks2007} 
A more systematic study of the low-bias conductance was carried out by B\"ohler 
\emph{et al.}\cite{Bohler2007} In this case, the authors conducted low-temperature 
(10 K) break junction experiments in which the molecules were evaporated \emph{in situ}. 
From the analysis of a conductance histogram, it was concluded that Au-C$_{60}$-Au 
junctions exhibit a preferred conductance value close $0.1G_0$. On the contrary, 
recent STM break junction experiments at room temperature have shown a very large 
spread of conductances with a certain preference for values around $5 \times 
10^{-4}G_{0}$.\cite{Yee2011} 

The main evidence that C$_{60}$ junctions can have a rather high conductance has been 
provided by a series of controlled STM experiments with other electrode materials (different
from gold) performed by Berndt and coworkers and which have been nicely backed up 
theoretically.\cite{Neel2007,Schull2009,Schull2011a,Schull2011b} Thus for instance, 
N\'eel \emph{et al.}\cite{Neel2007} reported a controlled STM study in ultrahigh 
vacuum (UHV) in which C$_{60}$ molecules deposited onto copper surfaces exhibited 
conductance values of the order of $0.25G_0$ in the contact regime. It is also
worth mentioning that Kiguchi\cite{Kiguchi2009} reported break junction experiments 
at room temperature in UHV conditions in which a single C$_{60}$ molecule between Pt 
electrodes was shown to exhibit a conductance as high as $0.7G_0$.

Although the electronic and transport properties of C$_{60}$ metal-molecule-metal 
junctions have been addressed theoretically by numerous groups,\cite{Neel2007,Schull2009,
Schull2011a,Schull2011b,Palacios2001a,Palacios2001b,Stadler2007,Shukla2008,Abad2010a,Abad2010b} 
studies of transport in C$_{60}$ junctions with gold electrodes, the most commonly used 
metal in molecular electronics, are surprisingly rather scarce.\cite{Ono2007,Zheng2009}
Moreover, those references explore only ideal contact geometries which may have
little to do with those realized in the experiments. For this reason, we present 
here an ab initio study of the charge transport in Au-C$_{60}$-Au junctions paying
special attention to the role of the contact geometry. Our analysis, based on the 
combination of density function theory (DFT) and nonequilibrium Green's function
techniques, shows that conductances above $0.1G_0$ are possible in realistic 
contact geometries. Moreover, motivated by very recent experiments,\cite{Yee2011} 
we have investigated the thermopower of these junctions and found that this quantity
is negative, as reported in Ref.~\onlinecite{Yee2011}, which is simply due to the fact 
that the low-bias transport is dominated by the C$_{60}$ LUMO. Furthermore, we have found 
that the thermopower varies significantly with the contact geometry, and its magnitude
can reach several tens of $\mu$V/K, which is higher than in previously investigated
molecules.\cite{Reddy2007,Baheti2008}

The second topic that we want to address in this work is the role of C$_{60}$ as an 
anchoring group. Some of the main challenges in the field of molecular electronics 
are related to the fabrication of single-molecule junctions with very well-defined 
transport properties and the ability to tune those properties at will. A strategy 
that is being pursued to achieve these goals is the use of suitable anchoring
groups to bind the molecules to the metallic electrodes. The thiol ($-$SH) group is 
the most commonly used anchoring group, specially when the electrodes are made of 
gold, because of their high covalent bond strength.\cite{Chen2006} However, the 
thiol group has been shown to lead to a large variety of binding 
geometries,\cite{Muller2006,Li2008,Burkle2012} which implies a large spread in the 
observed conductance values. Many different alternatives to the thiol group have 
been explored in recent years. For instance, Venkataraman and 
coworkers\cite{Venkataraman2008} introduced the amine group 
($-$NH$_2$) as an interesting possibility to obtain better defined values in the 
conductance histograms, which was attributed to a higher selectivity of amine-gold 
binding. Similar conclusions have been drawn in a recent analysis of
nitrile-terminated ($-$C$\equiv$N) biphenyls.\cite{Mishchenko2011} The list of 
anchoring groups explored in molecular junctions increases steadily and the search 
for the ``most" convenient group, leading to highly reproducible transport properties, 
has become one of the central issues in molecular electronics.\cite{Kiguchi2008,
Zotti2010,Venkataraman2010}

In this context, Martin \emph{et al.}\cite{Martin2008} put forward the 
interesting idea of using C$_{60}$ as a new anchoring group. The idea is that
C$_{60}$ offers a large contact area which, together with the high molecular symmetry,
may reduce the spread of the conductance values. Moreover, this fullerene
is known to strongly hybridize with metallic surfaces,\cite{Rogero2002} and,
as explained above, it has been shown in different STM and break junction 
experiments that it can sustain a rather high conductance.
Indeed, in Ref.~\onlinecite{Martin2008} the electrical characteristics of 
1,4-bis(fullero[c]pyrrolidin-1-yl)benzene (BDC60) (with C$_{60}$ anchor groups)
were studied using gold microfabricated break junctions, and it was found 
that the conductance histograms exhibited more pronounced peaks than those 
obtained with 1,4-benzenediamine and 1,4-benzenedithiol. More recently, Leary 
\emph{et al.}\cite{Leary2011} have shown that the use of C$_{60}$ as anchoring 
group facilitates enormously the characterization of single-molecule junctions
in STM experiments under ambient conditions and it allows to establish unambiguously
the conductance of the molecule under study.

These experimental results are promising, but it remains to be explored
whether the use of C$_{60}$ as a terminal group still allows, for instance, for the 
possibility to chemically tune the conductance by an appropriate functionalization 
of the molecular core, as it has been demonstrated with other anchoring 
groups.\cite{Venkataraman2006b,Venkataraman2007,Leary2007,Mishchenko2010,
Mishchenko2011} In other words, the main role of an anchoring group must be 
to provide the chemical link to the electrodes without modifying the essential 
properties of the molecular backbone. In this sense, it remains to be shown 
whether or not C$_{60}$ is too invasive to be used as an anchoring group. 

Besides the results for the pure C$_{60}$, we also present a study of the transport 
properties of molecular junctions based on C$_{60}$-terminated molecules: BDC60 and 
several derivatives.  Our DFT-based analysis aims at addressing two main questions: 
(i) Does C$_{60}$ reduce the spread in conductance values found with other anchoring 
groups? and (ii) Is C$_{60}$ too invasive to be used as a suitable anchoring group? 
Our results suggest that the conductance and thermopower of C$_{60}$-terminated 
molecules are still quite sensitive to the binding geometry and we expect a 
large spread of values in typical STM and break junction experiments. On the 
other hand, our results indicate that C$_{60}$ may reduce the electron communication 
between the molecular core and the metallic electrodes, leading to a reduction 
of the conductance. However, this reduction of the effective metal-molecule 
coupling and the fact that the frontier orbitals lie relatively close to the 
Fermi energy lead to a notable increase in the sensitivity of the junctions to the 
functionalization of the molecular backbone, as compared with thiol or amine groups.

The rest of the paper is organized as follows. In the next section we briefly 
describe the methodology employed to compute the transport properties of 
single-molecule junctions. Then, in Sec.\ \ref{sec-C60} we present a detailed 
analysis of the conductance and thermopower of Au-C$_{60}$-Au junctions. Sec.\ 
\ref{sec-dumbbell} is devoted to the analysis of junctions with BDC60 molecules
modified by the inclusion of several side groups. The results for the conductance 
and thermopower of these junctions are compared with those obtained using thiol 
and amine as anchoring groups. Finally, we summarize the main conclusions of 
this work in Sec.\ \ref{sec-conclusions}.

\section{Methodology}

Our main goal is to describe the transport properties of single-molecule
junctions based on C$_{60}$ molecules and C$_{60}$-terminated compounds. For 
this purpose, we employed the DFT-based transport method described in detail 
in Ref.~\onlinecite{Pauly2008}, which is built upon the quantum-chemistry 
code TURBOMOLE 6.1 \cite{Turbomole}. In this method, the first step is the
description of the electronic structure of the molecular junctions within
DFT. In all our calculations we used the BP86 functional\cite{BP86} and
the def-SVP basis set.\cite{Schafer1992} In order to construct the junction
geometries, we first relaxed the molecules in the gas phase. Then, the molecular
junctions were constructed by placing the relaxed molecules between two finite 
clusters of 20 (or 19) gold atoms and performing a new geometry optimization.
In this optimization, the molecule and the 4 (or 3) outermost gold atoms on 
each side were relaxed, while the other gold atoms were kept frozen. 
Subsequently, the size of the gold clusters was extended to around 63 atoms
on each side in order to describe the metal-molecule charge transfer and the
energy level alignment correctly. 

The final step in our method is to transform the information about the electronic 
structure of the junctions obtained within DFT into the different transport 
properties. This is done using nonequilibrium Green's function techniques, 
as described in detail in Ref.~\onlinecite{Pauly2008}. In the coherent 
transport regime, and following the spirit of the Landauer approach, the 
low-temperature linear conductance is given by $G= G_0 \tau(E_{\rm F}) = G_0 
\sum_i \tau_i(E_{\rm F})$, where $G_0=2e^2/h$ is the quantum of conductance, 
$\tau(E_{\rm F})$ is the junction transmission at the Fermi energy, 
$E_{\rm F}$, and $\{ \tau_i(E) \}$ are the transmission coefficients, \emph{i.e.}, 
the energy-dependent eigenvalues of the transmission matrix. The second transport 
property of interest in this work is the thermopower, which within the coherent 
transport regime is given by
\begin{equation}
 S=-\frac{K_{1}(T)}{eTK_{0}(T)} ,
\label{S-eq1}
\end{equation}
with $K_{n}(T)= \int{dE (E-\mu)^n \tau(E) [-\partial_{E}f(E,T)]}$, where $\mu$ 
is the electrochemical potential and $f(E,T)= [ 1+ \exp[(E-\mu)/k_{\rm B}T]]^{-1}$.
We shall compute the thermopower at room temperature ($T=300$ K), and in all the examples
discussed here one can still use the low-temperature expansion of Eq.~(\ref{S-eq1}), 
which is given by
\begin{equation}
S= - \frac{\pi^2 k^2_{\rm B}T}{3e} \frac{\tau^{\prime}(E_{\rm F})}
{\tau(E_{\rm F})} .
\end{equation}
Here, the prime denotes a derivative with respect to energy. Thus, the thermopower 
measures the logarithmic first derivative of the transmission function at $E =
E_{\rm F}$. The sign of this quantity carries information about the location of 
the Fermi energy within the gap of a molecular junction.\cite{Paulsson2003,Pauly2008b}
 
\section{Conductance and thermopower of gold-C$_{60}$-gold junctions}
\label{sec-C60}

\begin{figure}[t]
\begin{center}
\includegraphics[width=8.2cm]{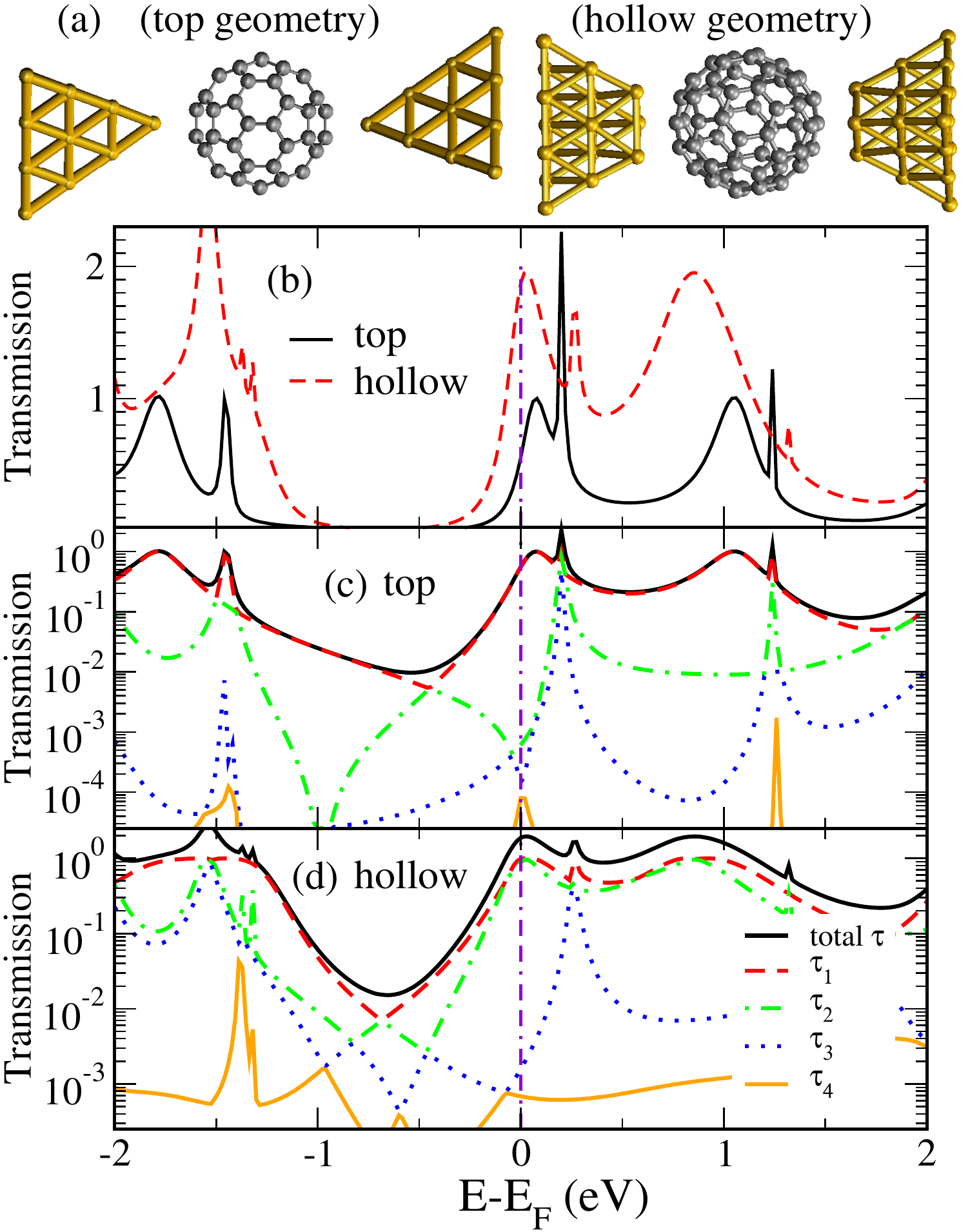}
\caption{(Color online) (a) Two ideal Au-C$_{60}$-Au junctions with top (left)
and hollow (right) binding geometries. (b) The transmission as a function of
energy for the two geometries in (a). (c-d) In these two panels the solid line
corresponds to the total transmission, while the others correspond to the
contribution of the individual transmission coefficients as a function of
energy.} \label{fig:top-hollow}
\end{center}
\end{figure}

This section is devoted to the analysis of the transport properties of Au-C$_{60}$-Au
junctions, which will also serve us as a reference for the study of the 
C$_{60}$-terminated molecules. Let us start our analysis by recalling the electronic 
structure of C$_{60}$ in the gas phase. Within our DFT approach, and in agreement with 
Ref.~\onlinecite{Ono2007}, we find that the highest occupied molecular orbital (HOMO) 
is five-fold degenerate, while the LUMO is three-fold degenerate 
(at $-5.90$ and $-4.26$ eV, respectively). These energies have to be compared with the 
Fermi energy of gold, which in our calculations is $-5.0$ eV. In order to elucidate 
how the electronic transport takes place through a C$_{60}$ molecule coupled to gold 
electrodes, we first consider two ideal geometries in which the molecule is bound 
to the electrodes in a top and in a hollow position, see Fig.~\ref{fig:top-hollow}(a). 
These geometries have been constructed as follows. We 
first relaxed the molecule on top of a single cluster, then we added a second 
cluster symmetrically at the other end, and finally we relaxed again the whole 
junction, as described in the previous section. In the top position, we find that 
the apex gold atom binds to two carbon atoms of a 6:6 bond, each C-Au distance 
being around 2.45~\AA{}. This geometry is consistent with that reported in 
Ref.~\onlinecite{Shukla2008} for various C$_{60}$-gold nanocontacts. In the 
hollow position, similar to that explored in Ref.~\onlinecite{Abad2010b}, 
the three-Au-atom terrace is facing one carbon atom and C-Au distances are 
in the range 2.3-2.4~\AA{}.

In Fig.~\ref{fig:top-hollow}(b-d) we show both the total 
transmission and its channel decomposition as a function of energy for these 
two ideal contact geometries. The first thing to notice is that the conductance,
determined by the transmission at the Fermi energy, is very high as compared
with other organic molecules of similar length, $0.55G_0$ and $1.85G_0$ for 
the top and hollow position, respectively. In both cases the low-bias conductance 
is dominated by the LUMO of the molecule, as it has been found in STM experiments 
of C$_{60}$ on Au surfaces (see e.g.\ Ref.~\onlinecite{Lu2004}). For the top 
position, we find that the transmission at the Fermi energy is largely dominated 
by a single channel which originates from one of the LUMOs of the molecule, 
which is split from the other two and is shifted to lower energies due
to its better coupling to the electrodes. For the hollow-type geometry, we 
find that two conduction channels give a significant contribution to the low-bias 
conductance. These channels originate from two of the LUMOs, which in this case 
are more strongly coupled to the electrodes than in the top geometry due to the 
higher number of C atoms in direct contact with the electrode atoms. This is simply 
the reason for the higher conductance of this geometry, which agrees with
the findings of Ref.~\onlinecite{Schull2011a}, where it was shown 
(both experimentally and theoretically) that the conductance of a C$_{60}$ 
junction increases with the number of atoms in contact with the molecule. 
Let us also mention that conductances above $1G_0$ have also been reported 
in theoretical studies of C$_{60}$-junctions with Al,\cite{Palacios2001a,Palacios2001b} 
Au,\cite{Ono2007} and Cu\cite{Schull2009} electrodes when the leads are similar
to ideal surfaces, \emph{i.e.}, with a high Au-C$_{60}$ coordination.

\begin{figure}[t]
\begin{center}
\includegraphics[width=8.8cm]{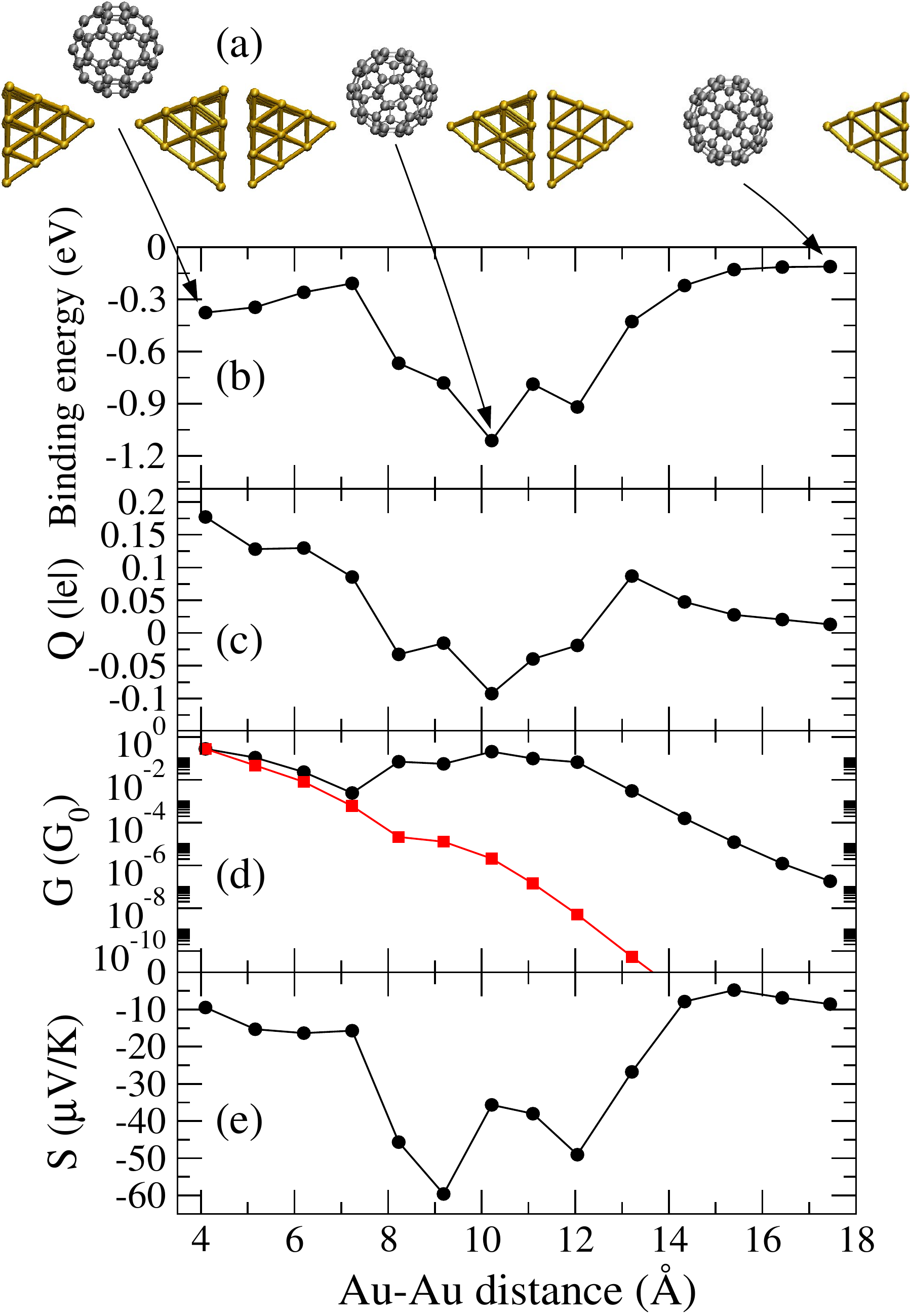}
\caption{(Color online) (a) Some representative geometries of the stretching simulation 
of a Au-C$_{60}$-Au junction. They correspond to Au-Au distances (distance between the 
Au tips) of 4.2 \AA{}, 10.2 \AA{}, and 17.5 \AA{}. The other panels show the following
quantities during the stretching process: (b) binding energy of the junction, (c)
charge on the C$_{60}$ molecule, (d) conductance of the Au-C$_{60}$-Au junction
(circles) and, for comparison, conductance of a Au-Au junction with the same
distance separation (squares), and (e) thermopower at room temperature.}
\label{fig:data-figures}
\end{center}
\end{figure}

The calculations discussed above suggest that Au-C$_{60}$-Au junctions can
have a conductance comparable to metallic atomic-size contacts. However,
the ideal geometries considered so far should provide a rough estimate for
the expected conductance values since it is unlikely to realize experimentally 
contacts with such a high degree of symmetry. Thus, a more direct comparison
with the experiments requires a detailed analysis of the junction formation
and of the evolution of the conductance during the stretching of the contacts.
This is precisely what we have done, as we now proceed to explain. In order
to simulate the junction formation, we started with a geometry in which the 
molecule is positioned laterally with respect to the gold-gold axis and we 
used gold clusters terminated with a single Au atom, Fig.~\ref{fig:data-figures}(a). 
Then, the gold electrodes were separated step-wise (in steps of $\sim 1$~\AA{}) 
and the junction geometry was relaxed in every step. This protocol was repeated 
until the junction was broken and the molecule lost contact with the 
electrodes. To characterize the junction during the stretching process,
we computed different quantities such as the binding energy of the junction,
the Mulliken charges in the C$_{60}$ molecule, the linear conductance, and 
the thermopower at room temperature. The results are shown in panels (b)
to (e) of Fig.~\ref{fig:data-figures}. Notice that in panel (d) we have also 
included the conductance of a Au-Au junction with the same distance separation
to estimate how much current is flowing directly from gold to gold (bypassing
the molecule) in the different stages of the elongation process.

In our simulation, after the first steps, the molecule rotates and then it
places itself in the middle of the junction adopting a geometry in which the top 
gold atom is bound to a single C atom, see central geometry of 
Fig.~\ref{fig:data-figures}(a). This structure, which is the most stable one with
a binding energy close to 1 eV, differs from the very symmetric top geometry in
Fig.~\ref{fig:top-hollow}(a). In this geometry the Au-C$_{60}$ interaction is 
maximized by the proximity of the side surface of the Au cluster. This is 
consistent with a related analysis reported in Ref.~\onlinecite{Stadler2007}. 
Then, after further stretching, the center of the molecule is aligned with the 
junction axis and the molecule remains there until the rupture of the contact. 
In the Au-Au distance range 8-12~\AA{}, where the binding energy is maximum in 
magnitude, the molecule is negatively charged and the conductance exhibits a 
``plateau" with values between 0.07 and $0.2G_0$, which is consistent with the 
experiments of Ref.~\onlinecite{Bohler2007}. At an Au-Au distance of $\sim$12~\AA{}, 
the contact breaks, as suggested by the evolution of the binding energy, and 
the conductance starts to decrease exponentially, while there is a tiny positive 
charge in the molecule.

\begin{figure}[t]
\begin{center}
\includegraphics[width=8cm]{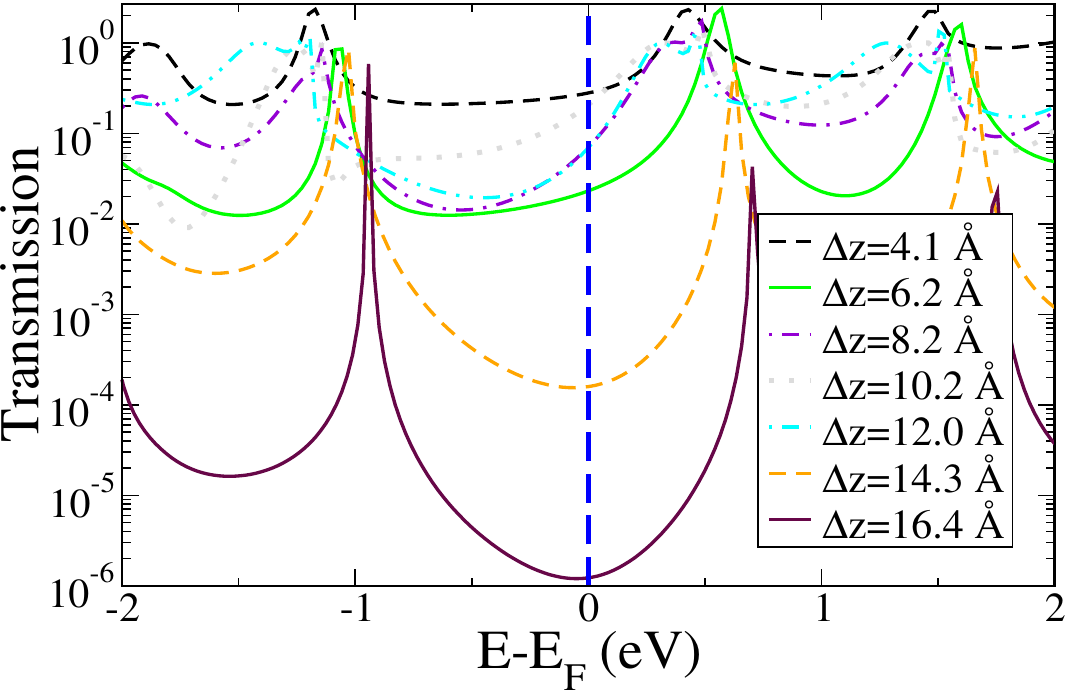}
\caption{(Color online) Transmission as a function of energy for 
the Au-C$_{60}$-Au junctions shown in Fig.~\ref{fig:stretch}(a). The
different curves correspond to different elongation stages, as indicated
in the legend ($\Delta z$ refers to the Au-Au distance).}
\label{fig:stretch}
\end{center}
\end{figure}

The thermopower results shown in Fig.~\ref{fig:data-figures}(e) deserve 
special attention in view of the recent experimental results reported in 
Ref.~\onlinecite{Yee2011}. In that work, thermopower measurements of
fullerene-metal junctions were performed at room temperature with a STM
break junction technique. In particular, for Au-C$_{60}$-Au junctions
a preferred value of $-14.5 \pm 1.2$~$\mu$V/K was found, the minus
sign suggesting that the LUMO dominates the conduction. As we show in
Fig.~\ref{fig:stretch}, where we display the transmission curves of 
Au-C$_{60}$-Au junctions at different elongation stages in the simulation,
the low-bias transport is dominated by the LUMO of the molecule at any 
distance. As a consequence, the thermopower is negative at any stage of
the elongation process, see Fig.~\ref{fig:data-figures}(e), and in particular,
its value for the most stable geometry is approximately $-35$~$\mu$V/K,
which is a factor two larger than in the experiment. Let us also mention
that for the ideal junctions of Fig.~\ref{fig:top-hollow}(a) we obtain
a thermopower of $-19.37$ $\mu$V/K for the hollow geometry and a value
of $-91.62$ $\mu$V/K for the top one. This big difference between these
two geometries is due to the fact that the transport takes place in an
almost on-resonant situation.

To conclude this section, it is worth commenting that in our simulations
we have not taken into account the van der Waals interactions. In this 
sense, the binding distances might not be exact. Dispersion forces play
a role in the interaction between C$_{60}$ and gold.\cite{Hamada2011}
However, the Au-C$_{60}$ binding is known to be mainly covalent with 
some ionic character.\cite{Hamada2011,Wang2004,Markussen2011}
This is corroborated by the fact that, during our simulated elongation,
the molecule is pulled in between the electrodes, due to the chemical
interaction.

\section{C$_{60}$ as an anchoring group} \label{sec-dumbbell}

\begin{figure}[b]
\begin{center}
\includegraphics[width=7.0cm]{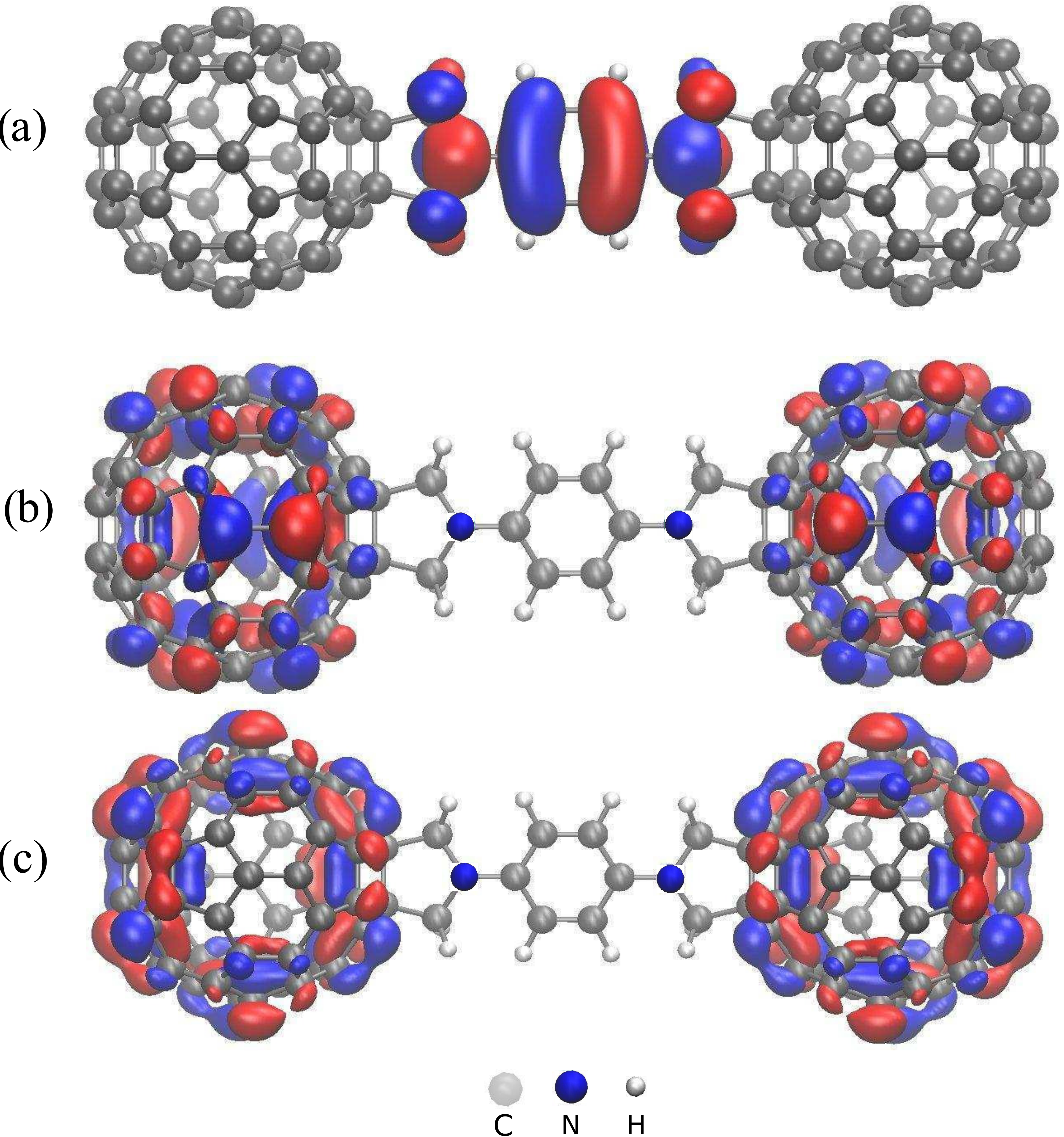}
\caption{(Color online) (a) HOMO and (b-c) two-fold degenerate LUMO
of the BDC60 in the gas phase.}
\label{BDC60-orbitals}
\end{center}
\end{figure}

We now analyze the role of C$_{60}$ as an anchoring group in
molecular junctions. We have seen in the previous section that this 
molecule can sustain a rather high conductance, which suggests that
C$_{60}$ can, in principle, provide a very efficient electronic 
communication, when used as an anchoring group. To explore this idea,
let us first study the transport through a 1,4-bis(fullero[c]pyrrolidinyl)benzene
(BDC60) molecule, see Fig.~\ref{BDC60-orbitals}, in which a phenyl ring is connected 
to two fullerenes on two opposite sides via a pyrrolidine group (in a so-called 
``dumbbell'' fashion). We have chosen this molecule for several reasons.
First, it has been investigated both experimentally\cite{Martin2008} and
theoretically,\cite{Geskin2011,Markussen2011} which allows us to establish
a comparison with our results. Second, the transport through the central 
moiety (a phenyl ring) can be analyzed with other anchoring groups, which
is necessary to determine the quality of C$_{60}$ as a terminal group. 

Our DFT calculations of the electronic structure of the isolated BDC60 molecule
show that its HOMO appears at -4.7 eV and that it is localized on the central part
of the molecule, while the two-fold degenerate LUMO is localized on the 
C$_{60}$'s, as displayed in Fig.~\ref{BDC60-orbitals}. These two LUMOs are 
the lowest in energy of a series of six levels (ranging from -4.18 to -3.87 
eV) which originate from the interaction between the three LUMO orbitals 
of each C$_{60}$, as explained in detail in Ref.~\onlinecite{Markussen2011}.
In agreement with this reference, we find that the nitrogen atoms are displaced 
from the phenyl ring plane. 

\begin{figure}[t]
\begin{center}
\includegraphics[width=8cm]{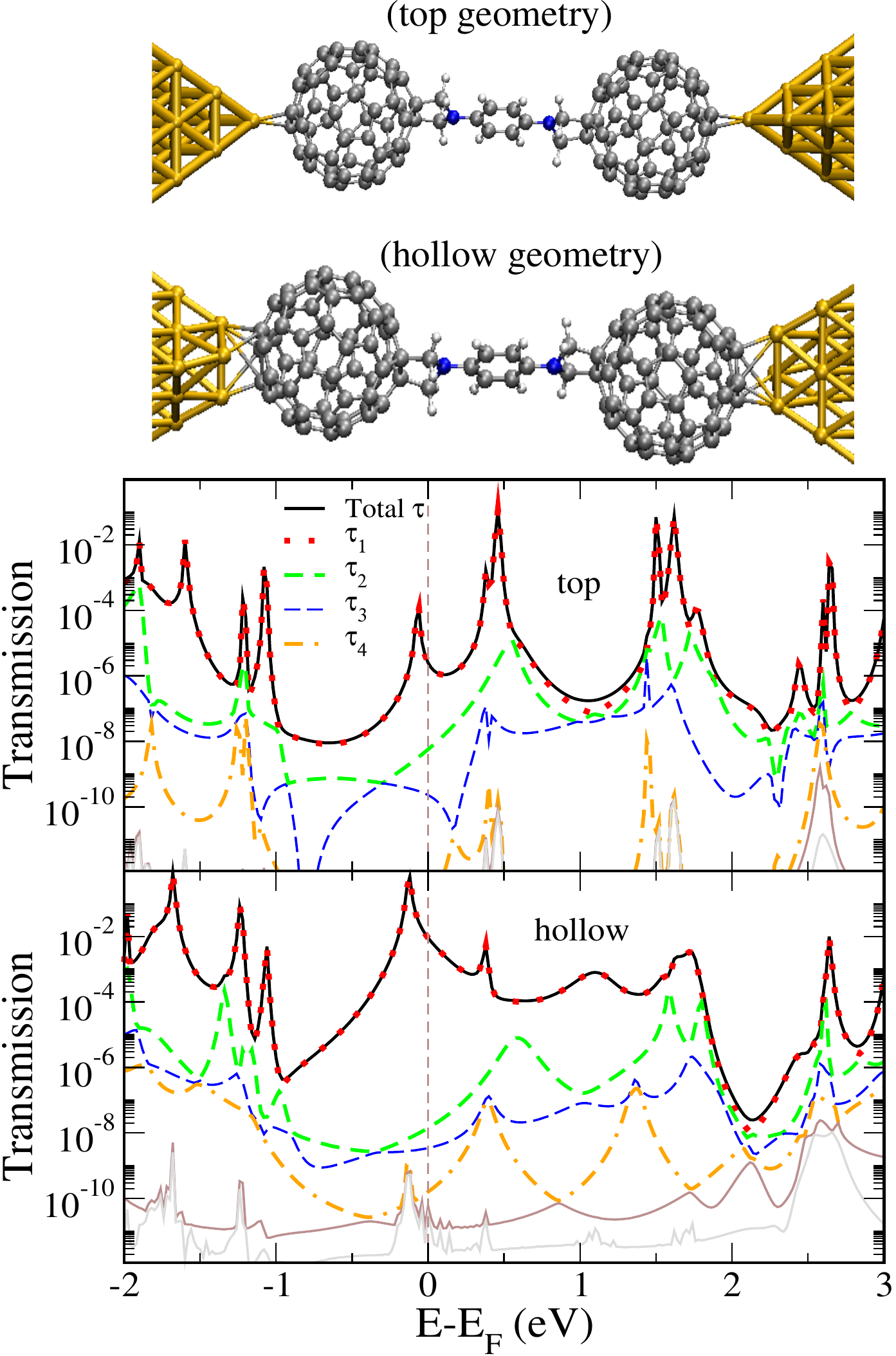}
\caption{(Color online) Total transmission and individual transmission coefficients
as a function of energy for the two Au-BDC60-Au junction geometries shown in
the upper part (top and hollow positions).}
\label{fig:paper-comp}
\end{center}
\end{figure}

Turning now to the analysis of the Au-BDC60-Au junctions,
we have again studied the conductance of two different types of geometries,
a hollow and a top binding geometry, see the upper part of Fig.~\ref{fig:paper-comp}. 
In the hollow position, the three top Au atoms are bound to four C atoms, 
while in the top position, the apex gold atom is bound to two C atoms of a 6:6 
bond. In Fig.~\ref{fig:paper-comp} we show the results for the total transmission
and the channel decomposition as a function of energy for both geometries. 
In both cases, the transmission close to the Fermi energy is dominated by a 
single channel and the resonance just below $E_{\rm F}$ originates from the
HOMO of the molecule. In spite of the fact that the HOMO is pinned very close to
the Fermi energy in both cases, the conductance is equal to $9.0 \times 10^{-3}
G_0$ for the hollow position, while it is $2.5 \times 10^{-6}G_0$ for the
top geometry. These values should be compared with the preferential value of 
$3 \times 10^{-4}G_0$ reported in the experiments of Ref.~\onlinecite{Martin2008},
although a large spread of conductance values was also found there.
We attribute the low values of the conductance (as compared to the C$_{60}$ 
junctions) and the difference between the two geometries to the weak effective 
coupling of the phenyl ring to the C$_{60}$ molecules, which is quite apparent 
in the small width of the transmission resonances. In other words, the
phenyl-C$_{60}$ effective coupling is the actual bottleneck in these junctions
and its weakness makes the conductance very sensitive to the exact level alignment
and to the metal-molecule coupling.

Let us compare our results for the BDC60 molecule with other theoretical results
published recently. First, we find that the current is mainly carried
by the HOMO of the BDC60, while in Ref.~\onlinecite{Markussen2011} it was
found that the transport is dominated by the LUMO. Let us stress that we have
confirmed the level alignment described above by test calculations with even
larger gold clusters (116 atoms). The discrepancy between these results may be
due to differences in the electrodes' shape (in Ref.~\onlinecite{Markussen2011}
the electrodes were modeled as ideal surfaces) and to the periodic boundary
conditions applied in their model.\cite{note1} Second,
in Ref.~\onlinecite{Markussen2011} it was claimed that the conductance
of the Au-BDC60-Au junctions is not very sensitive to the binding geometry,
while we find a large difference between the top and hollow geometries.
We attribute this discrepancy to the fact that in that reference no binding 
with undercoordinated Au atoms was considered. On the other hand, in
Ref.~\onlinecite{Geskin2011} it was stated that the weak coupling and the
insufficient conjugation throughout the three parts of this molecule is 
detrimental for the electronic transmission. Although our results
can not be directly compared with those of Ref.~\onlinecite{Geskin2011}
(supporting LUMO transport), we agree in observing that the peaks
corresponding to the frontier orbitals (HOMO and LUMO) appear
narrower and lower than in the transmission curve of C$_{60}$.

\begin{table}[t]
\begin{tabular}{ | c | c | c | c | c || c | c | c | c |}
\hline
 & \multicolumn{4}{c|| }{HOMO (eV)} & \multicolumn{4}{c|}{LUMO (eV)} \\
\cline{2-9}
    & CH$_3$ & H & F & Cl & CH$_3$ & H & F & Cl \\
\hline
S & -4.82& -4.96& -5.47&-5.46 &-1.23 &-1.42 & -1.75& -2.13 \\
NH$_2$ &-3.88 & -5.48& -5.92&-5.93 &-0.52 & -1.16&-1.67 & -1.97 \\
C$_{60}$ & -5.04& -4.70 & -5.24 & -5.56& -4.16& -4.18 & -4.19 & -4.18 \\
\hline
\end{tabular}
\caption{HOMO and LUMO energies of the molecules (in gas phase) based on
the phenyl unit and different anchoring and side groups.}
\label{tbl:1}
\end{table}

After studying BDC60, we want to address the issue of whether
or not the conductance of this dumbbell molecule can be chemically tuned by 
functionalizing the phenyl unit, as it is known to be possible with other
anchoring groups. To this aim, we have investigated three different 
substituents: CH$_{3}$, F, and Cl, and we shall compare the results with 
those obtained employing two other widely used anchoring groups, namely 
thiol ($-$SH) and amine ($-$NH$_2$). The CH$_3$ group is known to be 
electron-donating, while F and Cl are electron-withdrawing groups, and indeed 
for the molecules with thiol and amine anchoring groups the HOMO and LUMO 
are pushed upward in energy when the molecule is functionalized with CH$_3$, 
while they are pulled downward with F and Cl, see Table~\ref{tbl:1}. However, 
in the case of the dumbbell molecules this trend is not reproduced. The LUMO is
not affected by the presence of the side groups, as expected since it resides
in the C$_{60}$s, and the HOMO is shifted to lower energies also in the 
presence of the CH$_3$ group, see Table~\ref{tbl:1}. Moreover, in this case 
the functionalization causes a distortion of the central part, due to the 
interaction between the pyrrolydine and the substituents. This distortion, 
which does not occur in the case of SH and NH$_2$ because of their lower steric 
hindrance, is responsible for the unusual behavior of the CH$_3$ side group. 

\begin{figure}[t]
\begin{center}
\includegraphics[width=7cm]{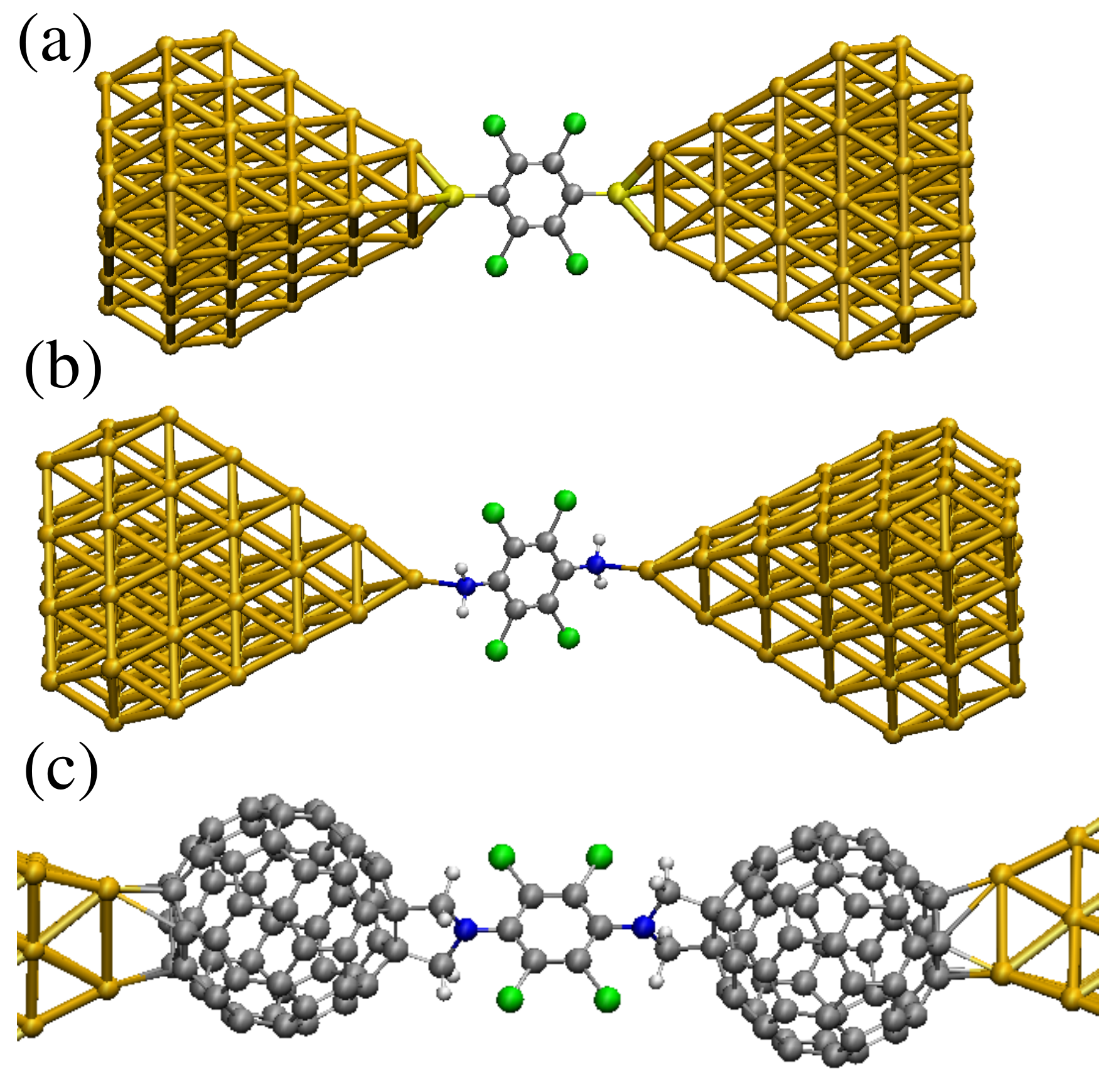}
\caption{(Color online) Geometries of the studied molecular junctions with a
phenyl ring functionalized with chlorine and attached to the gold electrodes
via (a) thiol, (b) amine, and (c) C$_{60}$ groups.}
\label{Cl-geometries}
\end{center}
\end{figure}

\begin{figure}[t]
\begin{center}
\includegraphics[width=8.5cm]{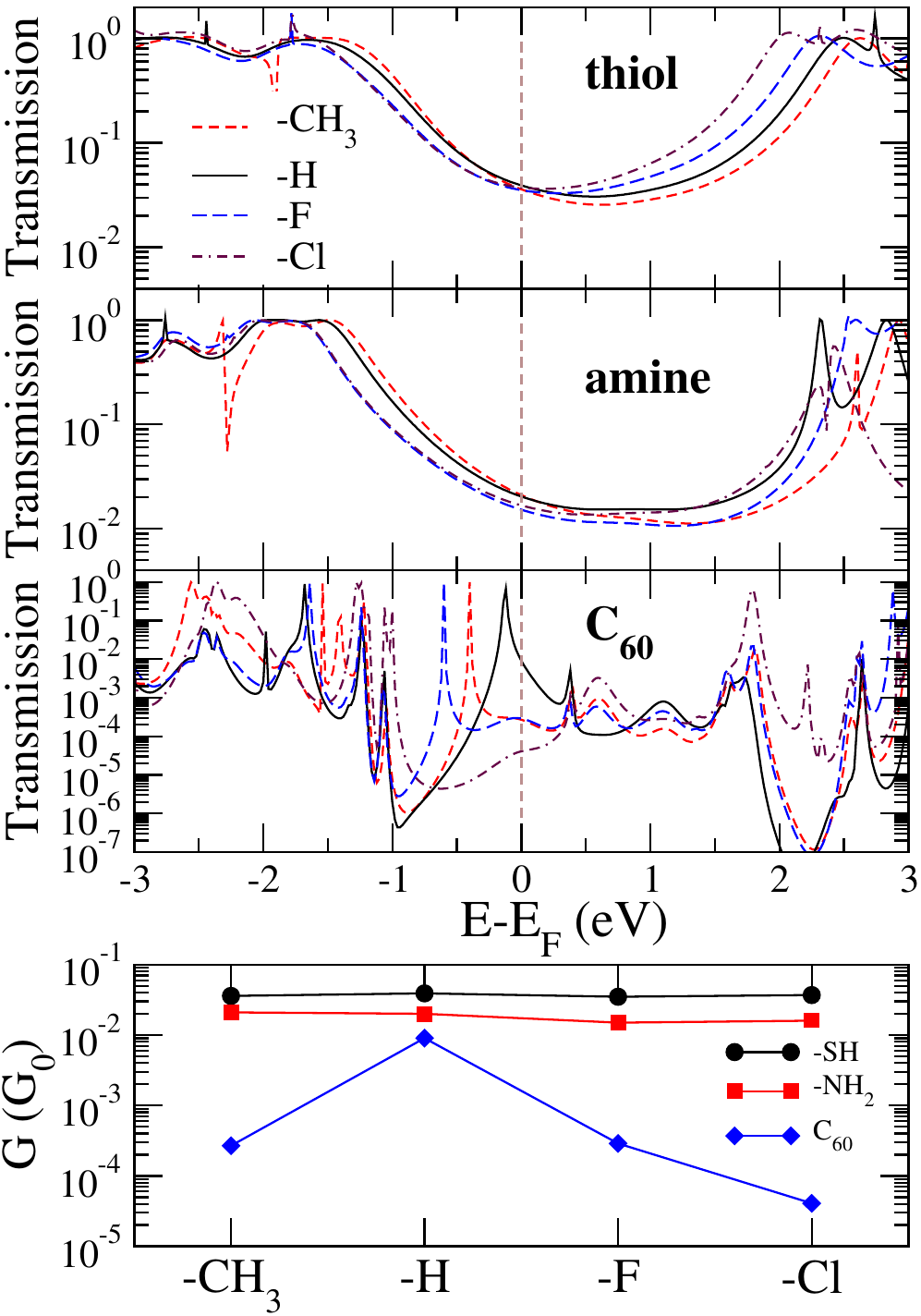}
\caption{(Color online) From top to bottom: Transmission curves for molecules
anchored via thiol, amine, and C$_{60}$ groups, and conductance values for all 
the junctions.} \label{fig:data}
\end{center}
\end{figure}

As an illustration, in Fig~\ref{Cl-geometries} we show the molecular junctions
for the three different anchoring groups and with Cl functionalization. 
We choose a top binding geometry for NH$_2$ (let us remind that the amine
group only binds to undercoordinated Au sites) and a hollow one for SH and 
C$_{60}$. In Fig~\ref{fig:data} we show the transmission curves for all 
molecules with different anchoring groups, as well as a comparison of 
the conductance values in the lowest panel. The first thing to notice is
that for all anchoring groups, the relative energy positions of the frontier
orbitals upon side functionalization reproduce the trends observed in the 
gas phase. For thiol- and amine-terminated molecules the current flows through 
the HOMO, consistent with what was found for tolane molecules with the same 
anchoring groups.\cite{Zotti2010} Concerning the conductance values, the 
functionalization has no dramatic effect in the cases of the thiol and amine 
group, see lower panel of Fig~\ref{fig:data}. The shift in the position of
the frontier orbitals is more apparent in the values of the thermopower,
which are shown in Table~\ref{tbl:2} for all the molecules. In this table
one can see the confirmation of the naive expectation that says that when 
the transport is dominated by the HOMO, the thermopower increases when
the HOMO is shifted to higher energies, and it decreases when this orbital
is pushed to lower energies. Moreover, it is worth stressing that the thermopower 
for the benzenedithiol molecules has been measured by Baheti \emph{et 
al.}\cite{Baheti2008} and our results are in good quantitative agreement.

\begin{table}[b]
\begin{tabular}{ | c | c | c | c | c |}
\hline
 & \multicolumn{4}{c|}{Thermopower ($\mu$V/K)} \\
\cline{2-5}
    & CH$_3$ & H & F & Cl \\
\hline
S & 8.36& 7.31& 5.18& 4.03 \\
NH$_2$ & 8.96 & 6.98&6.29& 5.64 \\
C$_{60}$ & 18.90 & 96.74 & 13.0 & -12.95 \\
\hline
\end{tabular}
\caption{Thermopower of the different molecular junctions analyzed in 
Fig.~\ref{fig:data} with different side groups and anchoring groups.}
\label{tbl:2}
\end{table}

As one can see in Fig~\ref{fig:data}, the effect of the side groups is much 
more pronounced in the case of the C$_{60}$-terminated molecules. In particular,
the functionalization in this case lowers considerably the conductance. The
main reason for that is that the transmission resonance that dominates the
transport, and which is associated to the HOMO of the molecule, is much narrower
for this anchoring group and therefore, it is much more sensitive to the 
shift induced by the side group. Notice also that in the case of the Cl group,
the conductance is indeed dominated by the LUMO of the molecule due to the
strong energy shift of the HOMO in this case. This fact is reflected in a 
change of sign in the thermopower (see Table~\ref{tbl:2}), which is something 
that does not occur for the other two anchoring groups.

\section{Conclusions} \label{sec-conclusions}

In summary, we have presented a DFT-based analysis of the conductance and
thermopower of individual C$_{60}$ and C$_{60}$-terminated molecules with
gold electrodes. We have shown, in agreement with several experiments, that
Au-C$_{60}$-Au junctions can have a rather high conductance above $0.1G_0$ 
for realistic geometries. Moreover, we have found that the transport through 
C$_{60}$ takes place through its LUMO, which leads to a negative thermopower 
in agreement with recent measurements. The fact that the LUMO lies 
relatively close to the Fermi energy, which means in practice that the energy 
derivative of the transmission at the Fermi energy is rather large, leads to 
a rather high thermopower in comparison with other organic molecules.

On the other hand, to investigate the use of C$_{60}$ as an anchoring group, 
we have first studied the transport through Au-BDC60-Au junctions and found 
that the conductance is rather sensitive to the binding geometry. Furthermore, 
we have found that the conductance is decreased, as compared with the C$_{60}$ 
junctions, due to the poor electronic communication between the C$_{60}$'s and 
the molecular core (phenyl unit). Then, in order to study whether C$_{60}$ as 
a terminal group is too invasive, we have analyzed several BDC60 derivatives 
which differ in the presence of a side group in the phenyl unit (Cl, F, and 
CH$_3$), and we have compared the results with those obtained using thiol and 
amine anchoring groups. Our results indicate that the BDC60-based junctions are 
much more sensitive to the functionalization, \emph{i.e.}\ the changes in the 
conductance and in the thermopower induced by the side groups are much more 
significant in the case of the molecules with C$_{60}$ as anchoring group. 

So in short, our study supports the idea that C$_{60}$ is a good conductor and
it suggests that it can be used as a convenient anchoring group to study typical 
effects related to the chemical modification of the molecules: role of side 
groups, degree of conjugation, length dependence, etc. However, C$_{60}$ does 
not seem to resolve the usual problem related to the spread of conductance values. 
Moreover, in dumbbell molecules like BDC60, beside the substituent-related shifting 
effect, also configurational changes due to steric repulsions can play an
important role.

\section{Acknowledgments}

We thank T. Frederiksen, G. Foti, E. Leary, and E. Scheer for fruitful discussions.
S.B, L.A.Z. and J.C.C. were funded by the EU through the network BIMORE
(MRTN-CT-2006-035859) and by the Comunidad de Madrid through the program 
NANOBIOMAGNET S2009/MAT1726. F.P. acknowledges funding through a Young 
Investigator Group and the DFG Center for Functional Nanostructures
(Project C3.6).


\end{document}